\documentclass[aps,prl,twocolumn,superscriptaddress,amsfonts,showpacs,floatfix]{revtex4}
\usepackage{amsmath,amssymb,bm,graphicx,epsfig,psfrag}

\newcommand\Ru{{\rm Re}}

\newcommand{\ii}{\imath}

\begin{document}

\title{Hindered energy cascade in highly helical isotropic turbulence}
\author{Rodion Stepanov}
\affiliation{Institute of Continuous Media Mechanics,  Korolyov 1,
Perm, 614013, Russia}
\affiliation{Perm National Research Polytechnic University, Komsomolskii av. 29, 614990 Perm, Russia}
\author{Ephim Golbraikh}
\affiliation{Ben-Gurion University, Physics Department, Beer-Sheva, Israel}
\author{Peter Frick}
\affiliation{Institute of Continuous Media Mechanics,  Korolyov 1,
Perm, 614013, Russia}
\author{Alexander Shestakov}
\affiliation{Institute of Continuous Media Mechanics,  Korolyov 1,
Perm, 614013, Russia}

\begin{abstract}
The conventional approach to the turbulent energy cascade, based on Richardson-Kolmogorov phenomenology, ignores the topology of emerging vortices, which is related to the helicity of the turbulent flow. It is generally believed that helicity can play a significant role in turbulent systems, e.g., supporting the generation of large-scale magnetic fields, but its impact on the energy cascade to small scales has never been observed. We suggest for the first time a generalized phenomenology for isotropic turbulence with an arbitrary spectral distribution of the helicity.
We discuss various scenarios of direct turbulent cascades with new helicity effect, which can be
interpreted as a hindering of the spectral energy transfer. Therefore the energy is accumulated and redistributed so that the efficiency of non-linear interactions will be sufficient to provide a constant energy flux.
We confirm our phenomenology by high Reynolds number numerical simulations based on a shell model of helical turbulence. The energy in our model is
injected at a certain large scale only, whereas the source of
helicity is distributed over all scales. In particular, we found that the helical bottleneck effect can appear in the inertial interval of the energy spectrum.

\end{abstract}


\maketitle

Many years ago, K.\,Moffat  delivered a verdict on the influence of
helicity on spectral properties of turbulent flows \cite{Moffat1978}:
\begin{quote}
\emph{no matter how strong the level of helicity injection may be
at wave-numbers of order $k_0$, the relative level of helicity as
measured by the dimensionless ratio $H(k) /2kE(k)$ must grow
progressively weaker with increasing $k$; and when $k/k_0$  is
sufficiently large it may be conjectured \cite{1973PhFl...16.1366B} that the helicity has negligible dynamical
effect, and is itself convected and diffused in much the same way
as a dynamically passive scalar contaminant \cite{1959JFM.....5..113B}}.
 \end{quote}
Thus the mean
helicity can play a significant role in turbulent systems, e.g., supporting the generation of large-scale magnetic fields \cite{Krause1980,Vainshtein1980}, but its impact on the spectral
properties of turbulent flows is practically negligible.

The experimental study of helicity needs hardly affordable three-dimensional measurements with high spacial resolution. Moreover, the helical forcing of a turbulent flow is not straightforward to implement in the laboratory. This is why the impact of helicity on the energy cascade has been studied in few laboratory experiments.  The energy and helicity spectra with a slope "-5/3"  have been measured in the atmospheric boundary layer \cite{2005DokPh..50..419K}.

Direct numerical simulations are able to model helical isotropic turbulence and evaluate the spectral properties of the flow.
They confirm the existence of Kolmogorov spectra (see for example
\cite{2003PhFl...15..361C,2008PhRvE..77d6303B}). The same results have been demonstrated at a high spectral resolution with the help of the shell model of turbulence \cite{2009FlDy...44..658S,2011PhFl...23c5108L}, showing  no
reason to doubt Moffat's  argument.

Features of helical turbulence are considered able to address the problem of the inverse energy cascade.
Like two-dimensional turbulence, it was suggested that two inertial ranges could be realized corresponding to the direct
cascade of helicity with an energy spectrum "-7/3" and an inverse cascade of the energy with a slope
"-5/3" \cite{1973PhFl...16.1366B, 1992PhFl....4..350W}. However, this analogy is not simply applicable because the helicity spectral density, in contrast to enstrophy, is not proportional to the energy spectral density.
Recently, direct numerical simulations of
the three-dimensional isotropic turbulence confirmed that the inverse cascade appears in
{\it truncated} three-dimensional Navier-Stokes equations \cite{PhysRevLett.108.164501}, in which the helicity is prescribed to be a sign-defined quantity and therefore directly related to the energy.

However, the statistical properties of helical turbulence are an issue. Indeed in some cases the spectra of  turbulent magnetohydrodynamical (see \cite{Branover1999,2015ExFl...56...88K} and references therein), convective \cite{2014PhFl...26f5103E} and atmospheric flows \cite{2006NPGeo..13..297G} differ from Kolmogorov's scaling, which can be interpreted as an effect of helicity. Furthermore, the influence of helicity on the flow properties can be manifested in the behavior of the turbulent spectrum of a passive scalar (e.g. an aerosol) \cite{1996JETP...83..192M}.  Helicity can change not only the energy and passive scalar spectra, but can lead to the
reduction of the turbulent viscosity and energy accumulation at large scales
\cite{1998PhyA..258...55B}, or can play a significant role in the generation of large-scale helical
vortex structures in the atmospheres of the Earth and planets
\cite{1983SPhD...28..926M,1986JAtS...43..126L,1996JETPL..63..813I}. To clarify the role of helicity, we consider the behavior of helical isotropic turbulence in a somewhat different way.

Coming back to forced helical turbulence, we revise the view that helicity is injected into the flow together with energy at the same scale. Theoretically, one can assume that turbulence is excited  by a source of energy at the largest scale and an independent source of pure helicity, acting at a certain scale or over all scales in the inertial interval.  Then, the helicity spectral flux is not constant anymore and the helicity spectral density can reach significantly higher values and influence the energy cascade. Real physical situations usually are far from ideal, but can be similar to some extent e.g. in rotating convective flows \cite{2013PhRvE..87c3016M,2013PhRvL.110l4501L}.

In the present work, we aim to elaborate on a phenomenology of an energy spectrum for the highly helical turbulence. Our theoretical result is justified by numerical simulations of helical turbulence with a controlled constant energy flux and distributed helicity injection over the scales in the inertial range. We demonstrate various scenarios of turbulent cascades with the helicity effect, and discuss its physical meaning with regard to the bottleneck phenomenon known for conventional turbulence \cite{1994PhFl....6.1411F}.

First, we adopt the basic statement of Kolmogorov's approach, which claims that in the inertial range at any scale $n$ the energy flux is equal to the dissipation rate, $\Pi_n^E=\varepsilon$. We consider a geometric sequence of scales and corresponding wave numbers $l_n^{-1}\sim k_n\sim  \lambda^n$. The energy flux is related to the velocity pulsations $v_n$ at this scale as
\begin{equation} \label{nonhelflux}
\Pi^E_n \approx  v_n^3 k_n
\end{equation}
and the energy of velocity pulsations $E_n \sim v_n^2\sim (\varepsilon / k_n)^{2/3}$.

Second, we follow the decomposition of velocity pulsations in two helical modes \cite{1992PhFl....4..350W}, $v_n = v_n^+ + v_n^-$, with corresponding energies
$E_n^{\pm}\sim (v_n^{\pm})^2$.
Then the energy and helicity at the scale $n$ are
\begin{eqnarray} \label{helmodes}
E_n=E_n^+ + E_n^-,
\nonumber \\
H_n=H_n^+ + H_n^-=k_n(E_n^+ - E_n^-).
\end{eqnarray}

The energy flux at scale $n$ is decomposed into four terms: $k_n(v_n^+)^2 v_n^-$, $k_n (v_n^-)^2 v_n^+$, $k_n(v_n^+)^3$  and
$k_n(v_n^-)^3$. 
The contribution of the first
two terms dominates in local interactions in the inertial range and becomes comparable
with the two other terms for non-local interaction \cite{1992PhFl....4..350W}. The significant role of terms
$k_n (v_n^\pm)^3$ was distinguished in the inverse cascade \cite{PhysRevLett.108.164501}. For the direct cascades of the energy and helicity one can neglect the contribution of these terms.

We introduce the relative helicity $H_n^r=H_n/(k_n E_n)$, which allows us to link the intensity of two helical modes
\begin{equation} \label{helmodes1}
v_n^-=v_n^+\sqrt\frac{1-H_n^r}{1+H_n^r} = \xi_n v_n^+.
\end{equation}
Then the energy flux provided by local interactions can be estimated as
\begin{equation} \label{helflux}
\Pi_n^E \approx k_n(v_n^+)^3 (\xi_n + \xi_n^2).
\end{equation}
Replacing (\ref{nonhelflux}) by (\ref{helflux}) we finally obtain
\begin{eqnarray} \label{helmodes2}
E_n^+ \sim \left( {\frac{\varepsilon}{k_n(\xi_n +\xi_n^2)}}\right)^{2/3}.
\end{eqnarray}
One can express  $E_n^- = \xi_n^2 E_n^+$ from (\ref{helmodes}) and obtain the total energy
\begin{equation}\label{entot}
E_n = E_n^+ + E_n^- \sim \left({\varepsilon \zeta_n/k_n}\right)^{2/3},
\end{equation}
where the dimensionless variable $\zeta_n$
\begin{equation}\label{zeta}
\zeta_n=\frac{(1+\xi_n^2)^{3/2}}{\xi_n + \xi_n^2}
\end{equation} depends on $H_r(k)$ and defines ``the degree of helical blocking'' of the spectral energy flux at a given scale. $|H_r(k)|$ characterizes the dominance of some helical modes over others with the opposite sign, i.e. it is the helical part of the energy. Then a new parameter $\delta(k)=1-|H_r(k)|$ corresponds to the non-helical part of the energy, which is free of helicity.
For the highly helical case $H_n^r \to \pm1$, formula (\ref{zeta}) has as asymptote
\begin{equation}\label{zeta2}
\zeta_n \approx \delta_n^{-1/2}.
\end{equation}
The corresponding spectral energy density
\begin{equation}\label{entot3}
E(k)  \approx \varepsilon ^{2/3} k^{-5/3} \delta(k)^{-1/3}
\end{equation}
is independent of the sign of the injected helicity.
Usually for the single-scale forcing of helicity, $H^r_n \propto k^{-1}$ and the parameter $\delta_n$ does not differ from unity. One can expect a significant change in the turbulent spectra for highly helical turbulence only.

We interpret our results as a hindering effect of the spectral energy transfer in scales with high relative helicity. The energy should be accumulated and redistributed so that the efficiency of non-linear interactions will be enough to provide the constant energy flux, which is predetermined by the energy injection rate. A similar consequence is observed as a result of the bottleneck phenomenon \cite{1994PhFl....6.1411F} in the non-helical turbulent cascade when non-local interactions drop out of the spectral energy transfer at the end of the inertial range. We exploit this analogy to name our effect the \emph{helical} bottleneck effect, having in mind the helical mechanism of cascade blocking.

\begin{figure*}
\centering
\hspace{5cm}\small(a)\hspace{5.4cm}\small(b)\hspace{5.4cm}\small(c)\vspace{-0.2cm}\\
\includegraphics[width=0.32\textwidth]{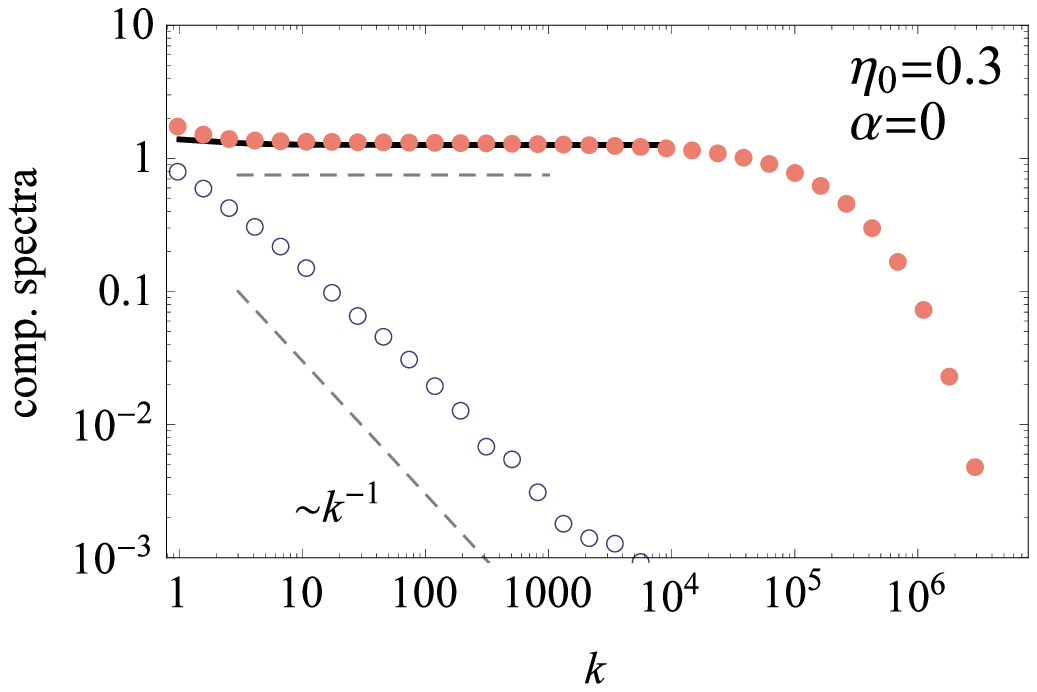}
\includegraphics[width=0.32\textwidth]{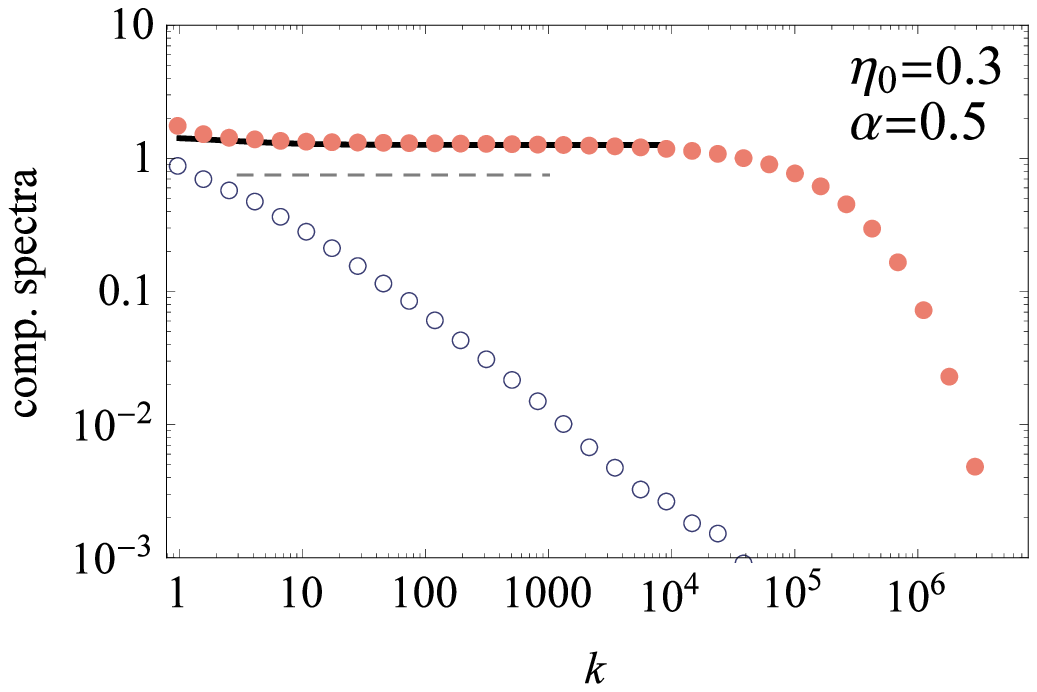}
\includegraphics[width=0.32\textwidth]{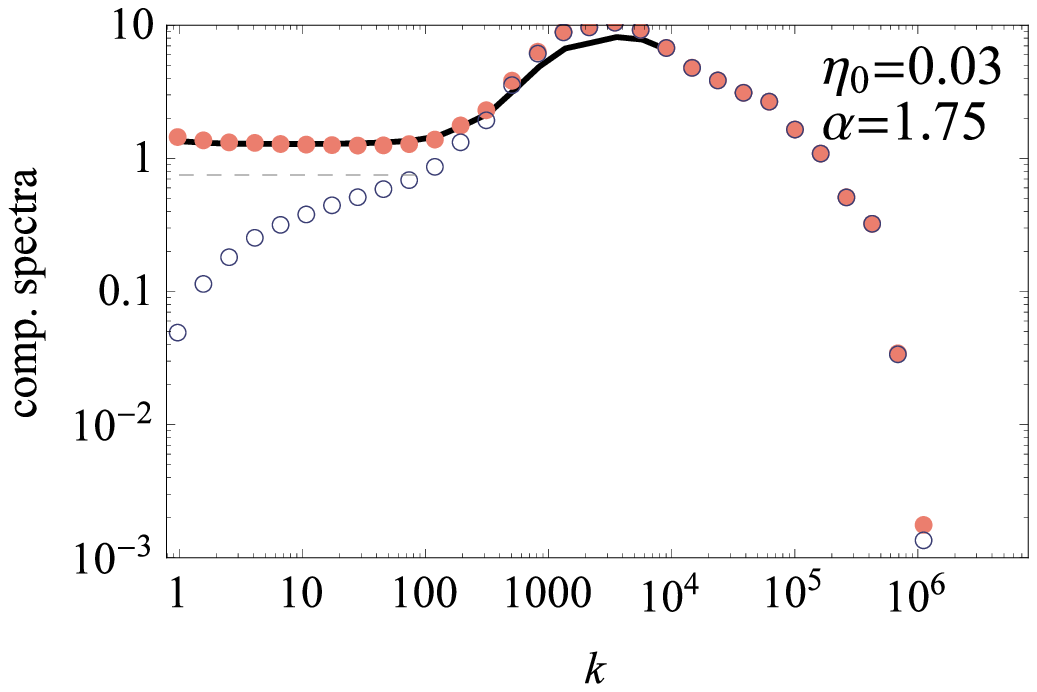}\\
\vspace{-0.3cm}
\hspace{5cm}\small(d)\hspace{5.4cm}\small(e)\hspace{5.4cm}\small(f)\vspace{-0.2cm}\\
\includegraphics[width=0.32\textwidth]{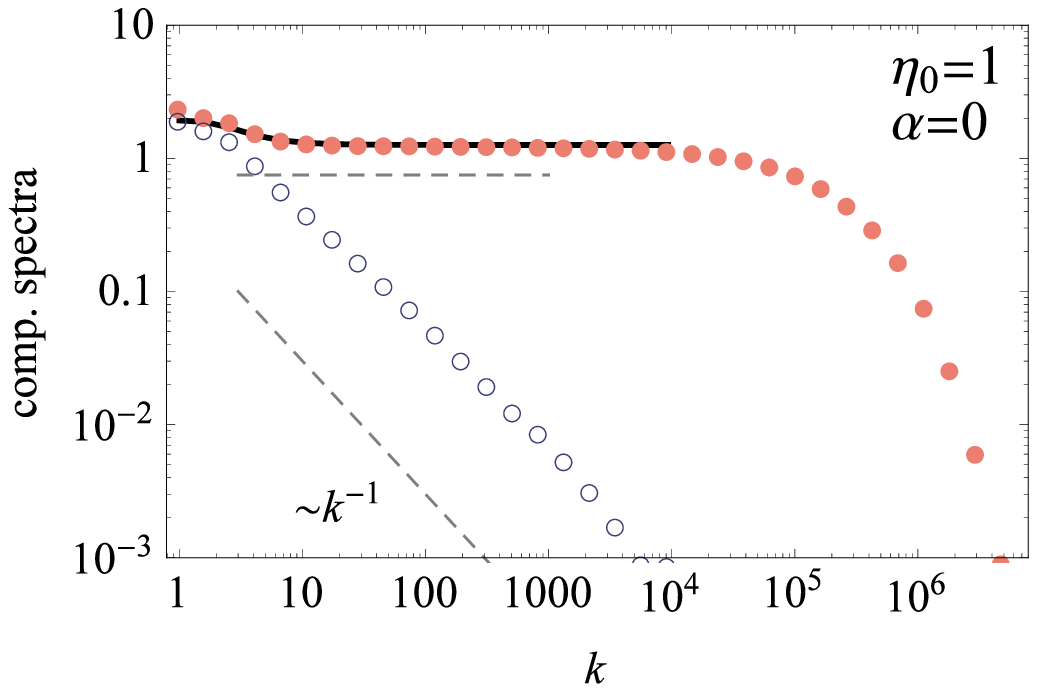}
\includegraphics[width=0.32\textwidth]{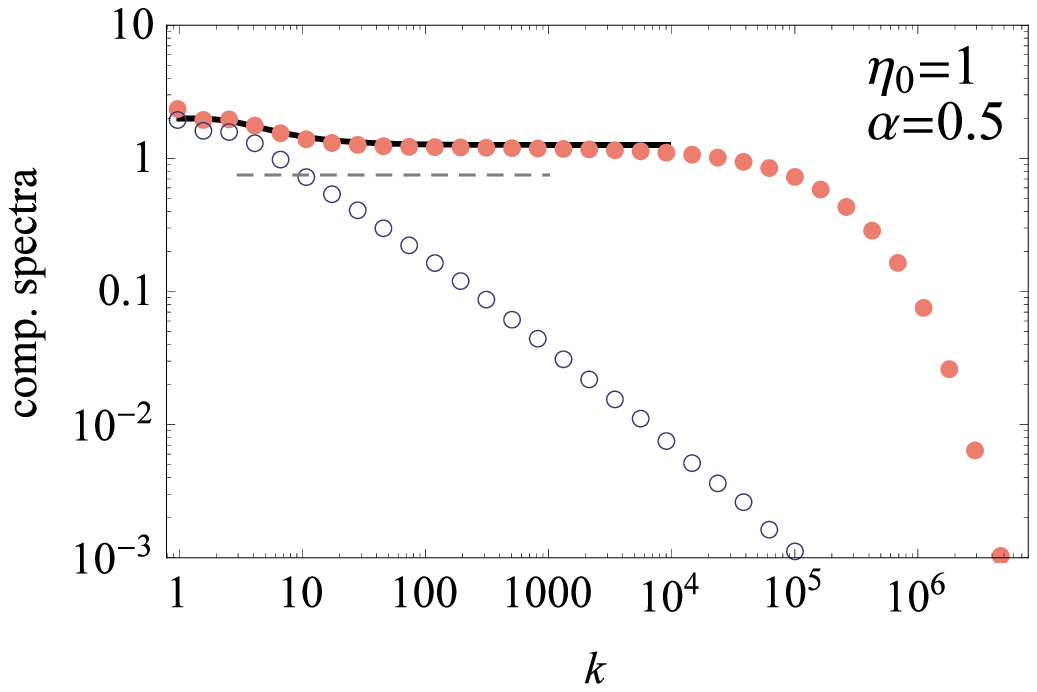}
\includegraphics[width=0.32\textwidth]{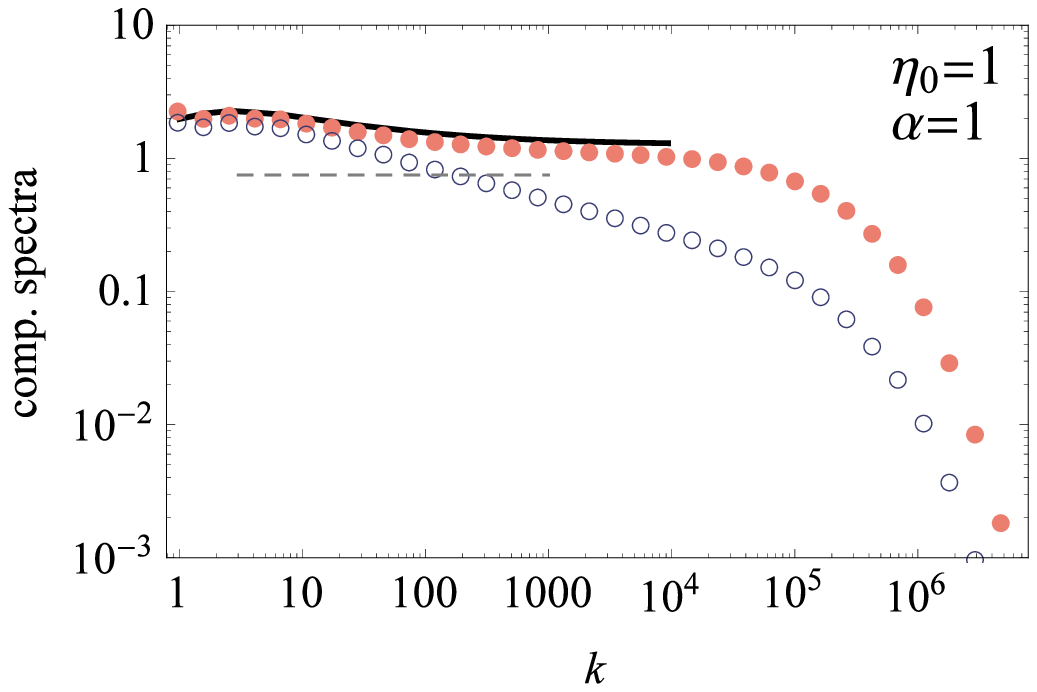}\\
\vspace{-0.3cm}
\hspace{5cm}\small(g)\hspace{5.4cm}\small(h)\hspace{5.4cm}\small(i)\vspace{-0.2cm}\\
\includegraphics[width=0.32\textwidth]{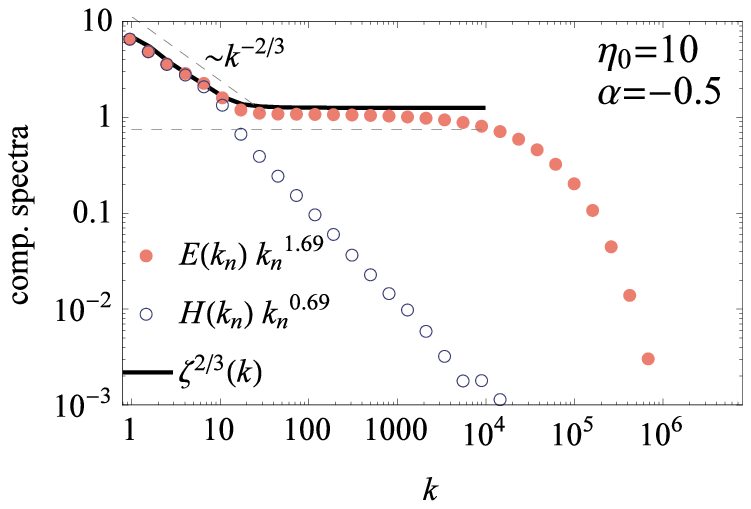}
\includegraphics[width=0.32\textwidth]{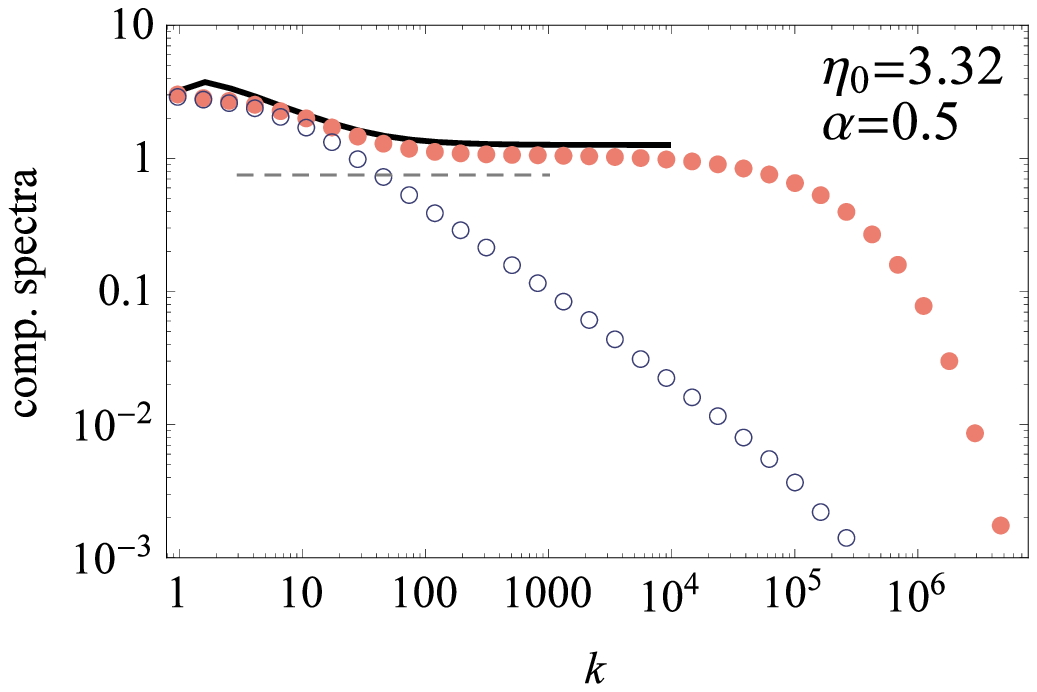}
\includegraphics[width=0.32\textwidth]{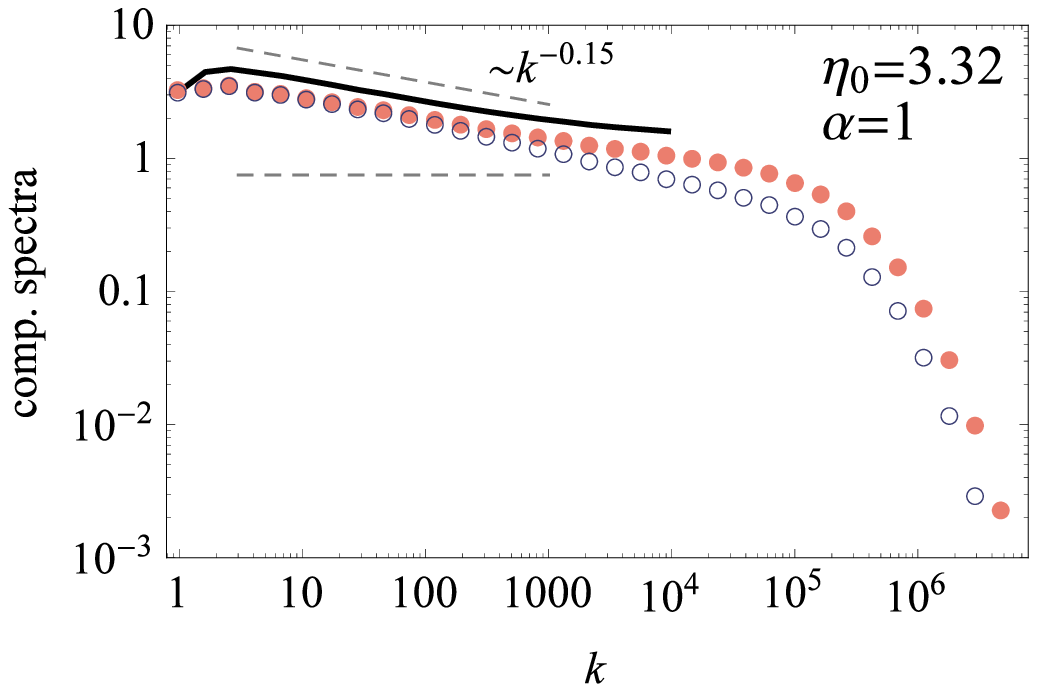}
\caption {Energy ($E(k)$, points) and normalized helicity ($H(k)/k$, circles) spectra  compensated by $k^{-1.69}$ (Kolmogorov's slope with intermittency correction) are shown in a grid of panels for each pairs of parameters
$\eta_0=0.3, 1, 3.32$ (from top to bottom) and $\alpha=0,0.5,1$ (from left to right). Exceptional cases are panel (c) $\eta_0=0.03, \alpha=1.75$ and panel (g) $\eta_0=10, \alpha=-0.5$. Theoretically predicted helical correction to the energy spectrum  $\zeta(k)$ (defined by Eq.~(\ref{zeta})) is shown by the solid line. } \label{fig1}
\end{figure*}

Further study of the suggested helical effect by DNS would be highly desirable.  Recent simulations  \cite{PhysRevE.92.031004} confirm a change of energy spectrum due to distributed helicity and justify a dominant contribution of triads $k_n (v_n^\pm)^2 v_n^\mp $ which is assumed in Eq.~(\ref{helflux}). However verification of Eq.~(\ref{entot3}) is hardly affordable by DNS with a resolution which seems to be achievable in near future. 
We demonstrate the realizability of the energy spectrum (\ref{entot}) with helical correction (\ref{zeta}) by numerical simulation using the helical shell model of turbulence (H1 type model as it is classified in the review \cite{2013PhR...523....1P}).  The governing equations for complex shell variables $U_n$ are
\begin{equation} \label{eq:shell}
\dot{U}_n= W_n(U,U)- {\Ru}^{-1} k^2_n U_n +f^E_n+f^H_n,
\end{equation}
where $k_n=\lambda^n$ is the wave number attributed to the shell
$n$, $\lambda$ characterizes the shell thickness, a bilinear operator $W_n$ determines interactions of $U_n$ with its neighbors $U_{n+1}$ and $U_{n-1}$.
The energy forcing term $f^E_n$ is prescribed to operate at the largest
scale only, i.e. in the shell $n=0$ ($k_0=1$), and
provides the mean energy injection rate $\varepsilon$ and the zero
mean helicity injection rate. For the helicity forcing, we introduce
a scale distributed force
\begin{equation}\label{forceH}
    f^H_n=\ii \eta_0 k_n^\alpha  U_n (U_n^2+{U_n^*}^2)/2,
\end{equation}
which acts in all shells and includes two parameters $\eta_0$ and $\alpha$ (the asterisk is for
complex conjugation).  $\eta_0$ defines the helicity
injection rate at the largest scale ($n=0$), which is the energy forcing scale also. $\alpha$ specifies the scaling of the helicity influx.

Variables $U_n$ define the energy $E_n=|U_n|^2/2$ and
helicity $H_n=\ii k_n
({U_n^*}^2-U_n^2)/4$ of the shell $n$. The corresponding spectra are $E(k_n)=E_n/k_n$ and $H(k_n)=H_n/k_n$.
As decomposition (\ref{helmodes}), one can split  $E_n$ and $H_n$  into two helical parts.
Then the helicity injection rate at the shell $n$ is
$\dot{H_n}= 4\eta_0 k_n^{\alpha+1} E_n^+ E_n^-$.
Note, that the force $f_n^H$ saturates and
tends to zero if the helicity $H_n$ approaches the limiting value $ k_n |U_n|^2$.

Numerical simulations of shell model equations (\ref{eq:shell}) are
performed for $\Ru=10^8$ and $\lambda=1.62$.
Statistically stationary spectra and fluxes are obtained by averaging over 256 runs (with random initial conditions and energy forcing), $10^4$ time units long each.
We choose the combination of governing parameters $\eta_0=0.3, 1, 3.32$ and $\alpha=0,0.5,1$  to study the transition of turbulence from low to high intensity of helicity injection (controlled by $\eta_0$) and from mostly large-scale injection to injection equally distributed over all scales (controlled by $\alpha$). Two special cases with $\eta_0=0.03$, $\alpha=1.75$ and  $\eta_0=10$, $\alpha=-0.5$ are included to emphasize the effect of helicity. The corresponding energy $E(k)$ and normalized helicity $H(k)/k$ spectra are presented in Fig.~\ref{fig1}. Both spectra are compensated by Kolmogorov's slope "-5/3" to highlight the difference in comparison with conventional turbulence. To get the flat energy and helicity spectra, we divide by a slightly steeper slope (caused by intermittency), namely, by $k^{-1.69}$. One can compare the result of simulations with the theoretically predicted energy spectrum (\ref{entot}), which is shown by a solid line for each set of the helical forcing parameters.

Panels (a) and (b) demonstrate the cases of weak helicity injection ($\eta_0=0.3$) in which passive helicity does not change energy spectrum. The helicity spectrum at $\alpha=0$ scales as $k^{-1}$, which looks like the spectrum for a passive scalar.  The moderate increase of $\alpha$ increases the slope of the helicity spectrum only (panel b).

Panels (d), (e) and (f) show results for stronger helicity injection ($\eta_0=1$) at which the helicity is becoming active.    Depending on $\alpha$, the energy accumulates in certain shells at large scales, i.e. in only four shells (panel d) or in all shells with $k_n<100$ (panel f).  In these ranges, the relative helicity tends to unity, which can be distinguished since $H(k)/k$ (open circles) approaches $E(k)$ (closed circles).

Panels (h) and (i) correspond to a further gradual intensification of helicity injection ($\eta_0=3.32$), which leads to a steeper slope of the energy spectrum. Again depending on $\alpha$, this can be observed in a part (panel h) or almost the entire inertial range (panel i). In the latter case, we find $E(k)\sim k^{-1.84}$ over three decades of scales.

Panel (g) presents a special case with the largest $\eta_0=10$ and lowest $\alpha=-0.5$. Two inertial ranges can be recognized with pronounced active and passive behavior of helicity.
A completely helical turbulent cascade, that is $E(k)\sim k^{- 7/3}$ and  $H(k)\sim k^{- 4/3}$, appears  at $k<10$. Kolmogorov's slope is recovered in the rest of the scales and there is a sharp border between the two ranges.

Panel (c) shows the opposite case with the lowest $\eta_0=0.03$ and largest $\alpha=1.75$. This combination of the forcing parameters provides a passive cascade of the helicity at the large scale and an active one at the small scale. This active range of  scales is characterized by the high relative helicity, which decelerates the energy cascade and results in the pileup of energy near the small-scale limit of the inertial range. This regime exaggerates the realistic scenario of small-scale helicity injection. However, it has particular interest because the result is similar to the bottleneck effect \cite{1994PhFl....6.1411F,Dobler2003}. Note, that spectra calculated using theoretical formula (\ref{zeta}), which are shown by solid lines in Fig.~\ref{fig1}, fit well the numerical spectra in all cases.

\begin{figure}
\centering
\vspace{0.2cm}
\hspace{5.5cm}\small(a)\vspace{-0.6cm}\\
\includegraphics[width=0.33\textwidth]{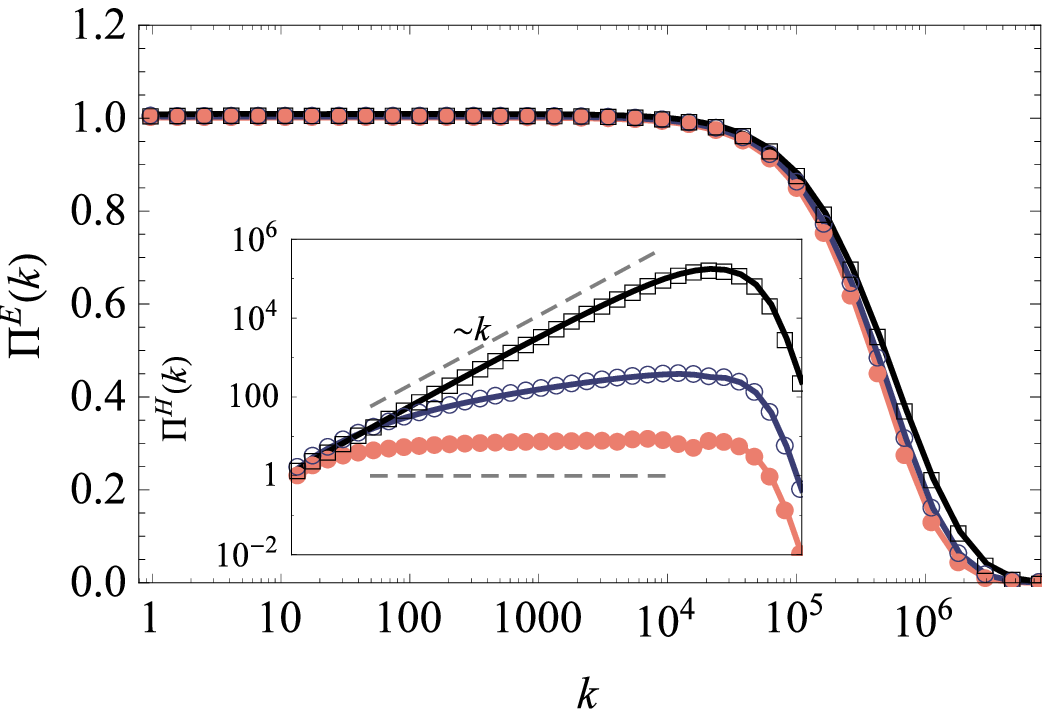}\\
\vspace{0.4cm}
\hspace{5.5cm}\small(b)\vspace{-1.cm}\\
\includegraphics[width=0.34\textwidth]{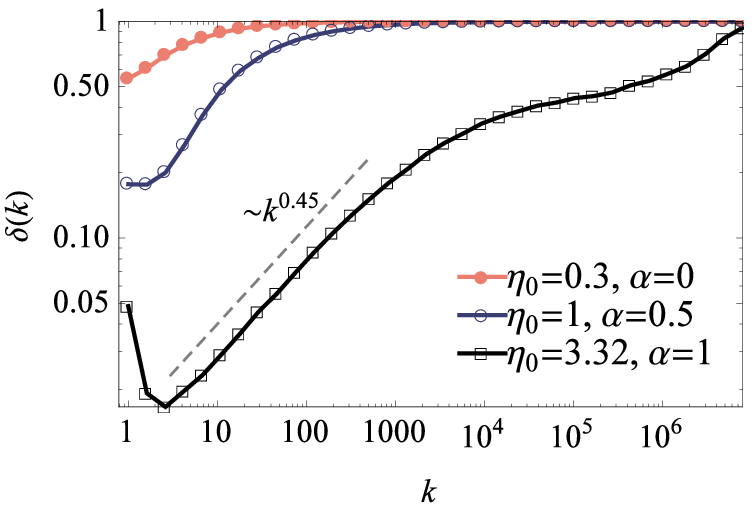}\\
\caption {(a) Spectral fluxes of energy and helicity (inset). (b) Distributions of $\delta(k)=1-|H_r(k)|$.} \label{fig2}
\end{figure}

The basic feature of the inertial range dynamics is the constant spectral energy flux.
Fig.~\ref{fig2}(a) presents the spectral fluxes of energy and helicity for three cases. The energy fluxes are not disturbed by the forcing (\ref{forceH}), even for a very strong injection of helicity. Apparently the helicity spectral flux is influenced by distributed helicity forcing. The helicity spectral flux continually approaches the asymptote $k$ with an increase of  $\eta_0$ and $\alpha$. $\Pi^H(k)\sim k \Pi^E(k)$ is the theoretical limit, which corresponds to the maximally helical turbulent cascade.

Instead of the relative helicity $H_r(k)$, the final expression (\ref{entot3}) includes $\delta(k)$. Both quantities $H_r(k)$ and $\delta(k)$ indicate scales where the energy is affected by the helicity ($H_r(k)\gtrsim0.7$ or $\delta(k)\lesssim0.3$). But $\delta(k)$  has a constant slope if the energy spectrum obeys a power law in some range of scales (see Fig.~\ref{fig2}(b)). Therefore, we conclude that the ''free'' part of the energy $\delta(k)$ is the more appropriate quantity which controls the efficiency of the spectral energy transfer.

Following Eq.~(\ref{entot3}), we should expect $\delta(k)\sim k^2$ if the power law "-7/3" appears in the energy spectrum. This is too steep to be realized over the whole inertial range. The limited range of scales with one order of magnitude is more realistic, as we find for the case in Fig.~\ref{fig1}(g). The affected range of scales shortens as the helical correction strengthens.  This is another argument in favour of the analogy with the bottleneck effect.

Our results expand the common point of view that the helicity is a possible trigger for inverse energy transfer, which support long-living large-scale vortical structures in astrophysical and  geophysical flows \cite{Levich1985}. We find a more modest role for helicity in reducing the efficiency of spectral energy transfer to small scales. There is no inverse cascade, just energy accumulation due to the hindered cascade in the highly helical flow. For example, we suggest that the helical bottleneck effect could be responsible for steepening the experimentally obtained energy spectrum in a von K{\'a}rm{\'a}n swirling flow \cite{2013PhRvL.110l4501L}. In that experiment, the helicity was intensively injected by the blades at smaller scales than the scale of energy injection by rotating discs \cite{Stefani}. Another applicable situation might be a flow with shear generated helicity, which is favorable for development and intensification of tropical cyclones \cite{1983SPhD...28..926M,2014JAtS...71.4308O}.  We note that the spacial imbalance of helicity can support a local intensification of vortical structures, even if the global helicity is negligible.

We revise the conventional approach to the turbulent energy cascade, based on Richardson-Kolmogorov phenomenology, where the topology of emerging vortices is ignored. This new generalization of the Kolmogorov phenomenology takes into account the arbitrary spectral distribution of helicity and describes a deviation from the -5/3 power law for the energy spectrum. The immediate consequence for the mean field theory is that an estimate of the effective turbulent diffusivity has to consider the helicity. This is an important message for physicists dealing with sub-grid models of turbulence.

\begin{acknowledgments}

This collaboration benefited from the International Research Group Program supported by Perm region Government.
Numerical simulations were performed on the supercomputers URAN and TRITON of Russian Academy of Science, Ural Branch.
\end{acknowledgments}

\bibliographystyle{apsrev}
\bibliography{ref}

\end{document}